\documentclass[aps,prl,twocolumn,superscriptaddress,showpacs,preprintnumbers,amsmath,amssymb]{revtex4-1}
\usepackage[colorlinks,linkcolor=blue,anchorcolor=blue,citecolor=blue]{hyperref}

\usepackage{graphicx} 
\usepackage{dcolumn}  
\usepackage{color}

\graphicspath{{ps}}

\newcommand{\ee}{e^{+}e^{-}}

\newcommand{\jp}{J/\psi}

\newcommand{\pipi}{\pi^{+}\pi^{-}}

\newcommand{\pim}{\pi^{-}}
\newcommand{\pip}{\pi^{+}}
\newcommand{\pipm}{\pi^{\pm}}
\newcommand{\piz}{\pi^{0}}

\newcommand{\rt}{\rightarrow}

\newcommand{\etal}{\em et al.}

\newcommand{\yns}{\Upsilon(nS)}
\newcommand{\zc}{Z_c}

\newcommand{\dstrz}{D^{*0}}
\newcommand{\dstrm}{D^{*-}}
\newcommand{\dstrp}{D^{*+}}
\newcommand{\dstrbar}{\bar{D}^{*}}
\newcommand{\dstrzbar}{\bar{D}^{*0}}
\newcommand{\dz}{D^{0}}

\newcommand{\dminus}{D^{-}}
\newcommand{\dplus}{D^{+}}

\newcommand{\done}{D_1}

\newcommand{\donebar}{\bar{D}_1}

\begin{document}



\title{ \quad\\[0.5cm] \boldmath Observation of a charged $(D\dstrbar)^{\pm}$ mass peak in
 $\ee\rt\pi D\dstrbar$ at $\sqrt{s}=4.26$~GeV}


\author{
\small
\begin{center}
M.~Ablikim$^{1}$, M.~N.~Achasov$^{8,a}$, O.~Albayrak$^{4}$, D.~J.~Ambrose$^{41}$, F.~F.~An$^{1}$, Q.~An$^{42}$, J.~Z.~Bai$^{1}$, R.~Baldini Ferroli$^{19A}$, Y.~Ban$^{28}$, J.~Becker$^{3}$
% ...
\\
\vspace{0.2cm}
(BESIII Collaboration)\\
\vspace{0.2cm} {\it
$^{1}$ Institute of High Energy Physics, Beijing 100049, People's Republic of China\\
% ...
$^{3}$ Bochum Ruhr-University, D-44780 Bochum, Germany\\
$^{4}$ Carnegie Mellon University, Pittsburgh, Pennsylvania 15213, USA\\
% ...
$^{8}$ G.I. Budker Institute of Nuclear Physics SB RAS (BINP), Novosibirsk 630090, Russia\\
% ...
$^{19}$ (A)INFN Laboratori Nazionali di Frascati, I-00044, Frascati, Italy; (B)INFN and University of Perugia, I-06100, Perugia, Italy\\
% ...
$^{28}$ Peking University, Beijing 100871, People's Republic of China\\
% ...
$^{41}$ University of Rochester, Rochester, New York 14627, USA\\
$^{42}$ University of Science and Technology of China, Hefei 230026, People's Republic of China\\
% ...
$^{a}$ Also at the Novosibirsk State University, Novosibirsk, 630090, Russia\\
% ...
}\end{center}
\vspace{0.4cm}
}

\begin{abstract}
We report on a study of the process $\ee\rt\pipm(D\dstrbar)^{\mp}$ at $\sqrt{s}=4.26$~GeV
using a 525 pb$^{-1}$ data sample collected with the BESIII detector at the BEPCII storage ring.
A distinct charged structure is observed in the $(D \dstrbar)^{\mp}$ invariant mass distribution.
When fitted to a mass-dependent-width Breit-Wigner lineshape, the pole mass and width are determined to be
$M_{\rm pole}=(3883.9 \pm 1.5 \pm 4.2)$~MeV/$c^2$ and $\Gamma_{\rm pole}=(24.8\pm 3.3 \pm 11.0)$~MeV.
The mass and width of the structure, which we refer to as $\zc(3885)$,
are $2\sigma$ and $1\sigma$, respectively, below those of the $Z_c(3900)\rt \pi^{\pm}\jp$ peak
observed by BESIII and Belle in $\pip\pim\jp$ final states produced at the
same center-of-mass energy. The angular distribution of the $\pi\zc(3885)$
system favors a $J^{P}=1^{+}$ quantum number assignment for the structure and disfavors $1^-$ or $0^-$.
The Born cross section times the $D\dstrbar$ branching fraction of the $\zc(3885)$ is measured
to be $\sigma(\ee \rt \pi^{\pm}\zc(3885)^{\mp})\times{\cal B}(\zc(3885)^{\mp}\rt(D\dstrbar)^{\mp})
=(83.5\pm6.6~\pm 22.0)$~pb. Assuming the $\zc(3885)\rt D\dstrbar$
signal reported here and the $Z_c(3900)\rt\pi\jp$ signal are from the same source,
the partial width ratio
$\frac{\Gamma(\zc(3885)\rt D\dstrbar)}{\Gamma(Z_c(3900)\rt\pi\jp)}=6.2\pm1.1~\pm2.7$
is determined.
\end{abstract}

\pacs{14.40.Rt, 13.25.Gv, 14.40.Pq}

\maketitle


{\renewcommand{\thefootnote}{\fnsymbol{footnote}}}
\setcounter{footnote}{0}

The $Y(4260)$ resonance
was first seen by BaBar as a peak in
the $\ee\rt \pipi\jp$ cross section as a function of $e^+ e^-$ center-of-mass (CM) energy~\cite{babar_y4260}.
It was subsequently confirmed by CLEO~\cite{cleo_y4260}
and Belle~\cite{belle_y4260}. Its production via the $e^+ e^-$ annihilation process
requires the quantum numbers of the $Y(4260)$ to be $J^{PC}=1^{--}$.
A peculiar feature is the absence
of any apparent corresponding structure in the
cross sections for $\ee\rt D^{(*)}\bar{D}^{(*)}(\pi)$ in the
$\sqrt{s}=4260$~MeV energy region~\cite{galina}. This implies
a lower-limit partial width of $\Gamma(Y(4260)\rt\pipi\jp)>1$~MeV~\cite{mo}
that is one order-of-magnitude larger than measured values for
conventional charmonium meson transitions~\cite{pdg}, and
indicates that the $Y(4260)$ is probably not a conventional
quarkonium state.

A similar pattern is seen in the $b$-quark sector, where
anomalously large cross sections for $\ee\rt\pipi\yns$
($n=1,2,3$) at energies around $\sqrt{s}=10.86$~GeV
reported by Belle~\cite{belle_pipiyns} were subsequently found to be associated
with the production of charged bottomonium-like resonances, the $Z_b(10610)^+$
and $Z_b(10650)^+$, both with strong decays to $\pi^+\Upsilon(nS)$ and
$\pi^+ h_b(mP)$ ($m=1,2$)~\cite{belle_z_b}.
The $Z_b(10610)^+$ mass is just above the $m_{B}+m_{B^*}$
threshold and it decays copiously to $B\bar{B}^*$,
while the $Z_b(10650)^+$ mass is just above the $2m_{B^*}$
threshold and it decays copiously to
$B^* \bar{B}^*$~\cite{belle_zb_bbstr}. Their proximity
to the $B\bar{B}^*$ and $B^*\bar{B}^*$ thresholds as well as
their decay patterns suggest that
these states may be molecule-like meson-meson virtual
states~\cite{voloshin}; a subject of considerable interest~\cite{molecule}.

Recently BESIII reported the observation of
a prominent resonance-like charged structure
in the $\pi\jp$ invariant mass distribution
for $\ee\rt\pipi\jp$ events collected at $\sqrt{s}=4.26$~GeV,
dubbed the $Z_c(3900)$.  A fit to a Breit-Wigner (BW) resonance lineshape
yields $M = (3899.0 \pm 3.6 \pm 4.9)$~MeV/$c^2$ and $\Gamma = (46 \pm 10 \pm 20)$~MeV~\cite{bes3_z3900}.
(Here, and elsewhere in this report, the first errors are
statistical and the second systematic.) This
observation was subsequently confirmed by
Belle~\cite{belle_z3900}. The $Z_c(3900)$ mass is $\sim$20~MeV/$c^2$ above the
$D\bar{D}^*$ mass threshold, which is suggestive of a virtual $D\bar{D}^*$
molecule-like structure~\cite{zhao,mahajan}; {\it i.e.}, a charmed-sector
analog of the $Z_b(1610)$.  (BESIII also reported resonance-like
structures in charged $D^*\bar{D}^*$ and $\pi h_c$ systems at $M\simeq 4025$~MeV, which
may be a charmed-sector analog of the $Z_b(10650$~\cite{bes3_z4025}.)
Another possibility is a diquark-diantiquark state~\cite{faccini}. It is important
to measure the rate for $Z_c(3900)$ decays to $D\bar{D}^*$ and
compare it to that of the $\pi\jp$.

Here we report the observation of a peak in the $(D\dstrbar)^-$ invariant-mass
distribution in $\ee\rt\pip (D\dstrbar)^{-}$ annihilation events
at $\sqrt{s}=4.26$~GeV with a 525~pb$^{-1}$ data sample detected by
the BESIII detector at the BEPCII electron-positron collider. In the following,
this structure is referred to as the $\zc(3885)$.
The $\pi^{+} (D\dstrbar)^{-}$ final states are selected by means of a
partial reconstruction technique in which only the bachelor
$\pi^{+}$ and one final-state $D$ meson are detected, and the presence of the
$\dstrbar$ is inferred from energy-momentum conservation.
(In this report,
the inclusion of charge conjugate states is always implied.)
We perform parallel analyses of both isospin channels
($\pip \dz {D}^{*-}$ and $\pim \dplus \dstrzbar)$ as a consistency check.
The $D$ mesons are reconstructed in the
$\dz\rt K^-\pi^+$ and $\dplus\rt K^-\pip\pip$ decay channels.

The BESIII detector is a large-solid-angle magnetic
spectrometer consisting of
a 50-layer Helium-gas-based main cylindrical drift chamber (MDC),
a barrel-like arrangement of time-of-flight  scintillation
counters (TOF), and an electromagnetic calorimeter
comprised of CsI(Tl) crystals  located inside
a superconducting solenoid coil that provides a 1~T
magnetic field.  An iron flux-return located outside of
the coil is instrumented with resistive plate chambers to
identify muons. The charged particle momentum resolution
for 1~GeV/$c$ charged tracks is
0.5\% and the energy resolution for 1~GeV photons is 2.5\%.
Measurements of $dE/dx$ in the MDC and flight times in the TOF
are combined to determine pion, kaon and proton identification (ID)
probabilities. The hypothesis with the highest ID probability
is assigned to each particle. The detector is described in detail
in Ref.~\cite{BESIII}.

To study the detector response and identify potential backgrounds,
we use samples of Monte Carlo (MC) simulated events that are produced
by the EVTGEN generator~\cite{evtgen} in conjunction with
KKMC~\cite{kkmc}, which generates ISR photons,
and simulated using a GEANT4-based~\cite{geant4} software package~\cite{boost}.
In addition to signal channels and various
potential background processes, we simulated generic events using Born cross sections for
charmonium processes that have been measured, Lundcharm to generate production of
other, non-measured charmonium states~\cite{lund}
and PYTHIA for unmeasured hadronic final states~\cite{pythia}.

For the $\pip\dz$-tag analysis, we select events with three
or more well reconstructed charged tracks in the polar angle region $|\cos\theta|< 0.93$,
with points of closest approach to the $\ee$ interaction point that are
less than $10$~cm in the beam direction
and 1~cm in the plane perpendicular to the beam direction.
At least one of the tracks is required to be negatively charged and
identified as a kaon. In addition, we require at least two positively charged tracks
that are identified as $\pip$ mesons.
We designate $K^-\pip$ combinations with invariant mass within $15$~MeV/$c^2$
of $m_{\dz}$ as $\dz$ candidates. For events with two or more $K^-\pip$ combinations,
we retain the one with invariant mass closest to $m_{\dz}$. For the $\pim\dplus$-tag
analysis, the selection is the same except for the requirement of an additional $\pim$
track that is identified as the bachelor pion and the mass requirement
$|M(K^-\pip\pip)-m_{\dplus}|<15$~MeV/$c^2$ to select the $\dplus$ candidates.

\begin{figure}[htb]
\begin{minipage}[t]{42mm}
  \includegraphics[height=0.8\textwidth,width=1.04\textwidth]{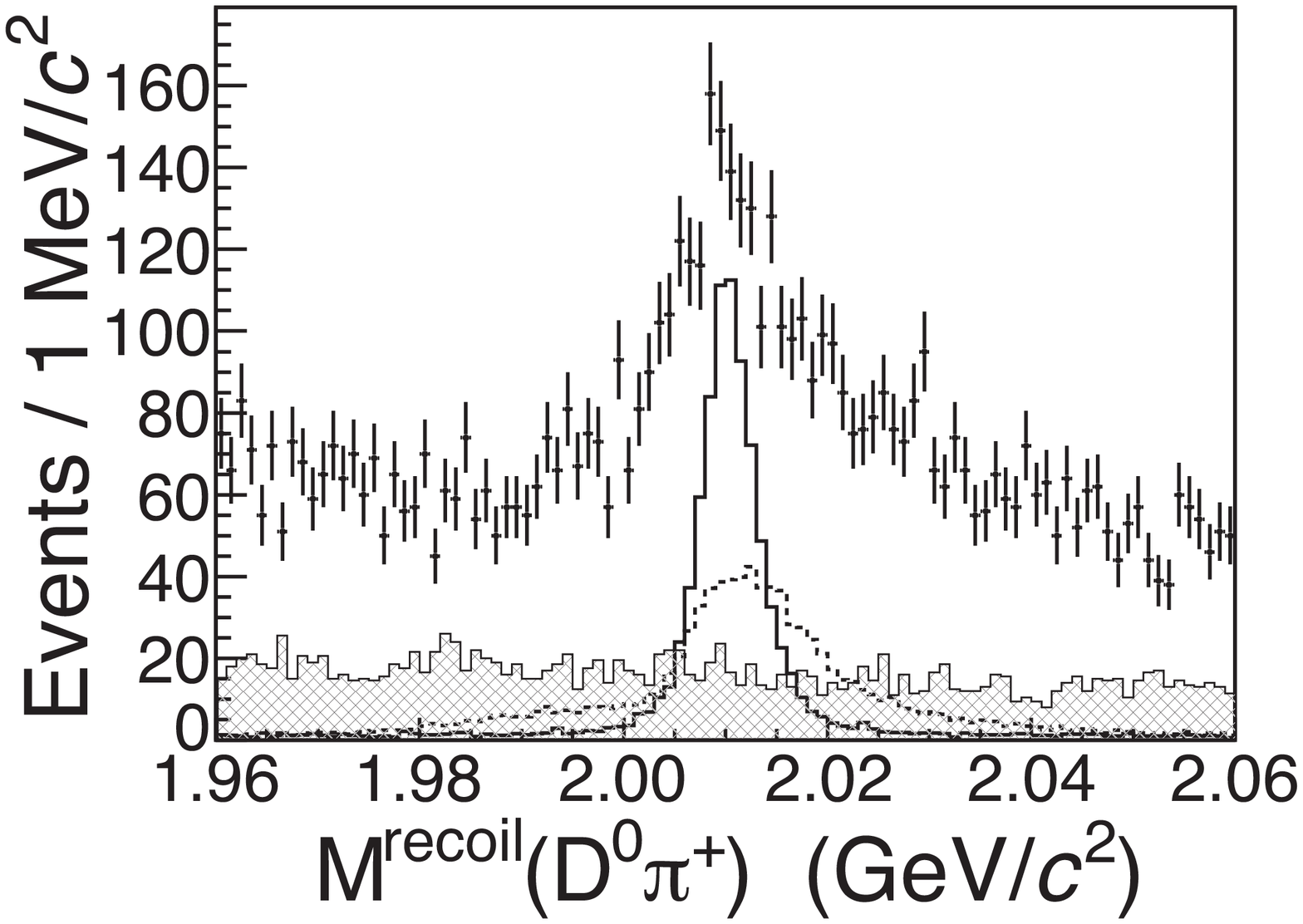}
\end{minipage}
\begin{minipage}[t]{42mm}
  \includegraphics[height=0.8\textwidth,width=1.04\textwidth]{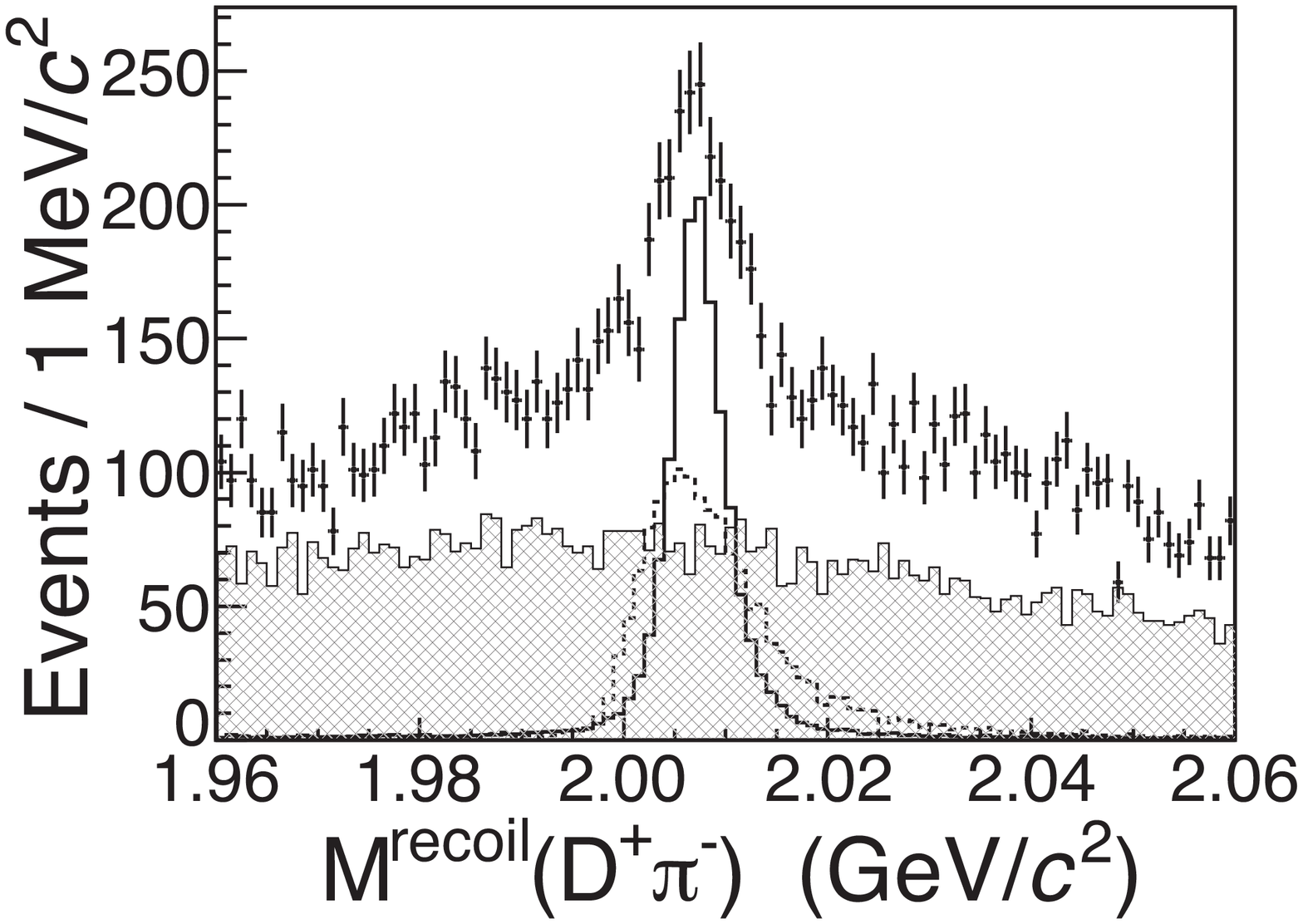}
\end{minipage}
\caption{\footnotesize The $\pi D$ recoil mass distribution for the
$\pip D^0$-  {\it (left)} and  $\pim D^+$-tagged {\it (right)} events. Points with errors
are data, the hatched histogram shows the events from the $D$ mass sidebands.
The solid and dashed histograms are described in the text.
}
\label{dst}
\end{figure}

The left panel of Fig.~\ref{dst} shows the distribution of masses recoiling
against the detected $\pip D^0$ system~\cite{recoil}, where a prominent peak at
$m_{D^{*-}}$ is evident.  The solid-line histogram shows the same distribution
for  MC-simulated $e^+e^-\rt \pip D^0 D^{*-}$,  $D^0\rt K^-\pip$ three-body phase-space
events.  Because of the limited phase space, some events from the isospin partner
decay $\pip\zc(3885)^-$, $\zc(3885)^-\rt \dminus\dstrz$, where  the detected $\dz$ is a decay product
of the $\dstrz$, also peak near $m_{D^{*-}}$, as shown by the dashed histogram that
is for MC-simulated $e^+e^-\rt \pip \zc(3885)^- $, $\zc(3885)^-\rt D^- D^{*0}$, $D^{*0}\rt \gamma$~or~$\pi^0 D^0$
decays.  Here the mass and width of the $\zc(3885)$ are set to our final measured values.
Since the $D\dstrbar$ invariant mass distribution is equivalent to the bachelor
pion recoil mass spectrum, the shape of the $\zc(3885)\rt D\bar{D}^*$ signal peak is
not sensitive to the parentage of the $D$ meson that is used for the event tagging.
The right panel of Fig.~\ref{dst} shows the corresponding plots for the $\pim D^+$ tagged events,
where the solid histogram shows the contribution from MC-simulated
$e^+e^-\rt \pim D^+ \bar{D}^{*0}$ three-body phase-space events.  Here, also,
the $\pim\dplus$-tagged event sample that is used to study $\pim\dplus\dstrzbar$ includes
some cross feed from  the $\pim\zc(3885)^+$, $\zc(3885)^+\rt\bar{D}^0\dstrp$ signal channel, where the
$\dplus$ used for tagging is a decay product of the $\dstrp$.  The dashed histogram is from
MC-simulated $e^+e^-\rt \pim \zc(3885)^+$, $ \zc(3885)^+\rt \bar{D}^0 D^{*+}$, $D^{*+}\rt \pi^0 D^+$ events.

\begin{figure}[htb]
\begin{minipage}[t]{42mm}
  \includegraphics[height=0.8\textwidth,width=1.04\textwidth]{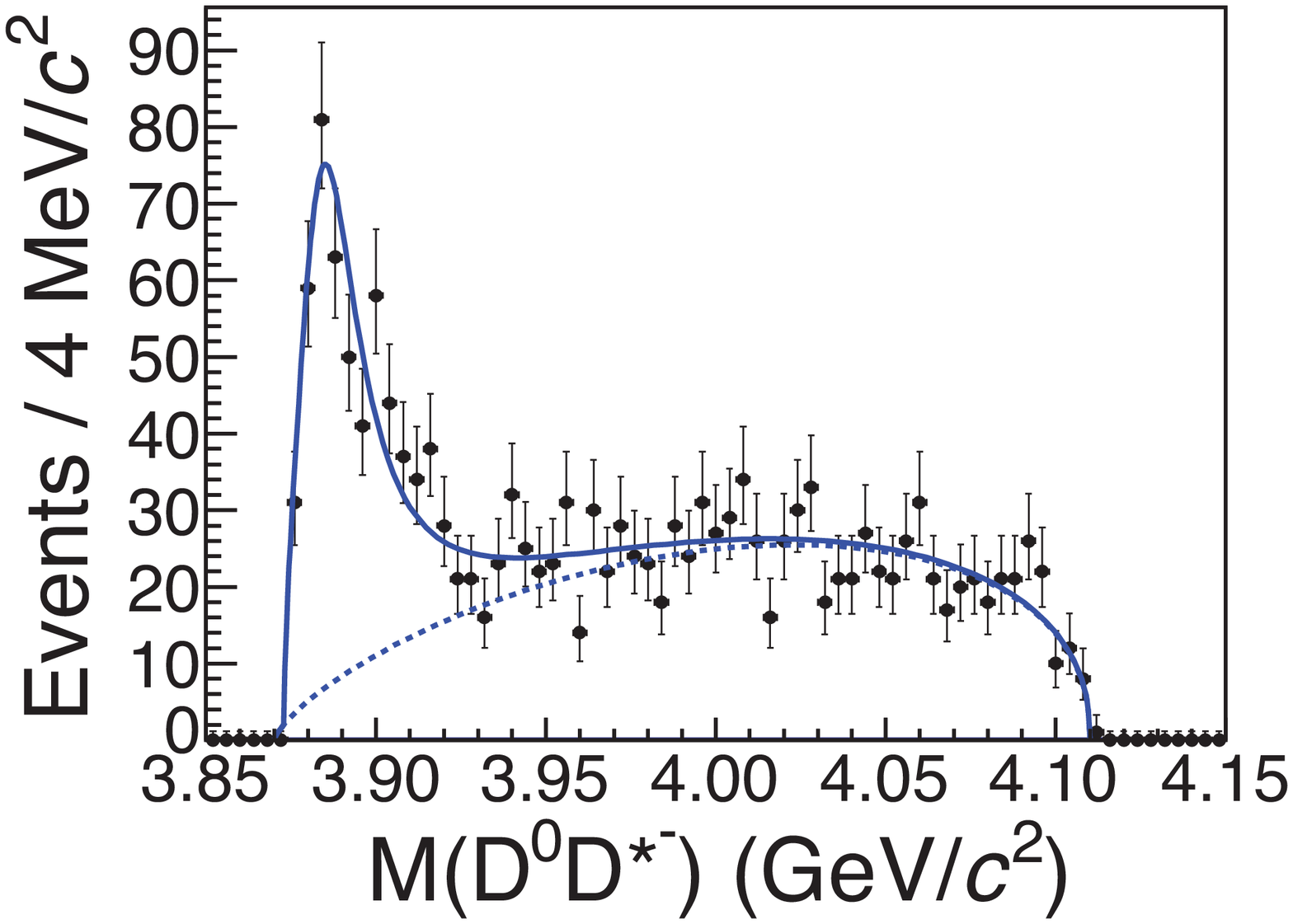}
\end{minipage}
\begin{minipage}[t]{42mm}
  \includegraphics[height=0.8\textwidth,width=1.04\textwidth]{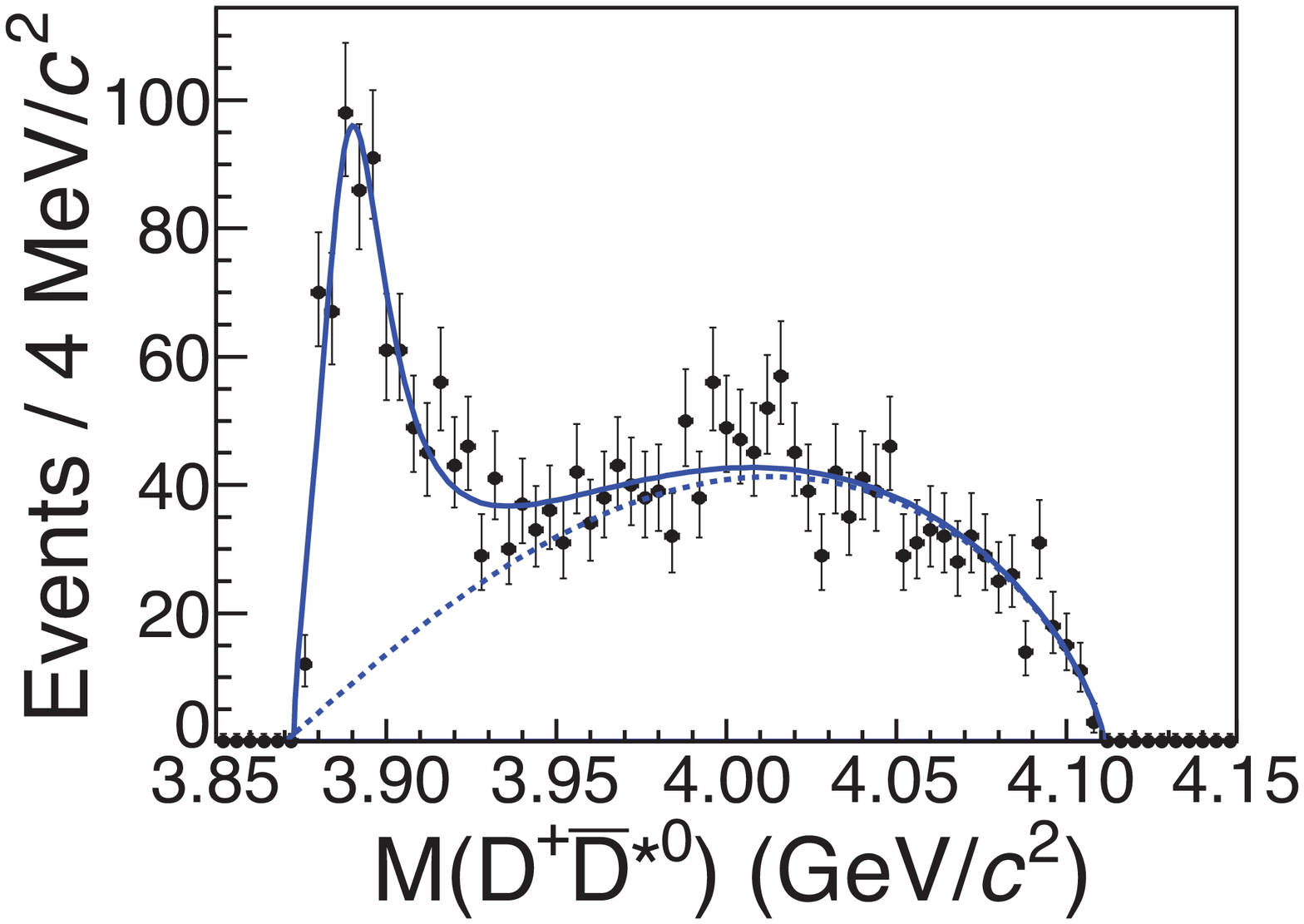}
\end{minipage}
\caption{\footnotesize The $M(\dz\dstrm)$ {\it (left)} and $M(\dplus\dstrzbar)$ {\it (right)}
distributions for selected events.  The curves are described in the text.
}
\label{fig:mddstr_2c}
\end{figure}

We apply a two-constraint kinematic fit to the selected events, where
we constrain the invariant mass of the $\dz$ ($\dplus$) candidate tracks
to be equal to $m_{\dz}$ ($m_{\dplus}$) and the mass recoiling from the $\pip\dz$
($\pim\dplus$) to be equal to $m_{\dstrm}$ ($m_{\dstrzbar}$). If there is more than
one bachelor pion candidate in an event, we retain the one with the smallest
$\chi^2$ from the kinematic fit. Events with $\chi^2<30$ are selected for further analysis.
For the $\pip\dz$-tag analysis, we require $M(\pip\dz)> 2.02$~GeV to reject the events of the type
$\ee\rt\dstrp\dstrm$, $\dstrp\rt\pip \dz$.
The left (right) panel of Fig.~\ref{fig:mddstr_2c} shows the distribution of $\dz\dstrm$ 
($D^+\dstrzbar$) invariant masses recoiling from the bachelor pion
for the $\pip\dz$ ($\pim\dplus$) tagged events.
The two distributions are similar and both have a distinct peak near the
$m_{D}+m_{\dstrbar}$ mass threshold.
{\color{black} For cross-feed events, the reconstructed $D$ meson is not in fact recoiling
from a $\dstrbar$ and the efficiency for satisfying these selection requirements decreases
with increasing $D\dstrbar$ mass. Studies with phase-space MC event samples show
that this acceptance variation is not sufficient to produce a peaking structure.}


To characterize the observed enhancement
and determine the signal yield,
we fit the histograms in the left and right panels of
Fig.~\ref{fig:mddstr_2c} using a mass-dependent-width Breit-Wigner (BW) lineshape
to model the signal and smooth threshold functions to
represent the non-peaking background. For the signal, we use
$dN/dm_{D\dstrbar} \propto (k^*)^{2\ell+1}|BW_{\zc}(m_{D\dstrbar})|^2$,
where $k^*$ is the $\zc$ momentum in the $\ee$ rest frame, $\ell$ is the
$\pi$-$\zc$ relative orbital angular momentum and
$BW_{\zc}(m_{D\dstrbar}) \propto \frac{\sqrt{m_{D\dstrbar}\Gamma_{\zc}}}
{m^2_{\zc} - m^2_{D\dstrbar}-im_{\zc}\Gamma_{\zc}}$. Here $\Gamma_{\zc}=\Gamma_{0}(q^*/q_{0})^{2L+1}(m_{\zc}/m_{D\dstrbar})$,
where $q^*(m_{D\dstrbar})$ is the $D$ momentum in the $\zc(3885)$ rest frame,
$q_0=q^*(m_{\zc})$ and $L$ is the $D$-$\dstrbar$ orbital angular momentum.
In the default fits, we set $\ell=0$, $L=0$ and leave $m_{\zc}$ and $\Gamma_{0}$ as free parameters.
We multiply the BW by a polynomial determined from a fit to the
MC-determined mass-dependent efficiency to form the signal
probability density function (PDF). Mass resolution effects are less than 1~MeV/$c^2$
and, thus, ignored. For the non-peaking background for the $M(D \dstrbar)$ distribution,
we use: $f_{\rm bkg}(m_{D\dstrbar})\propto (m_{D\dstrbar}-M_{\rm min})^c(M_{\rm max}-m_{D\dstrbar})^d$,
where $M_{\rm min}$ and $M_{\rm max}$ are the minimum and maximum kinematically allowed masses,
respectively. The exponents $c$ and $d$ are free parameters determined from the fits to the data.

The results of the fits are shown as solid curves in Fig.~\ref{fig:mddstr_2c}.
The dashed curves show the fitted non-resonant background. The fitted BW masses and widths from the $\pip D^0$
($\pim D^+$) tagged sample are $3889.2 \pm 1.8$~MeV/$c^2$ and $ 28.1 \pm 4.1 $~MeV
($3891.8 \pm 1.8$~MeV/$c^2$ and $ 27.8 \pm 3.9 $~MeV), where the errors are statistical only.
Since the mass and width of a mass-dependent-width BW are model dependent and may differ
from the actual resonance properties~\cite{pole},
we solve for $P=M_{\rm pole}-i\Gamma_{\rm pole}/2$,
the position in the complex $(M,\Gamma)$ plane where the BW denominator is zero,
and use $M_{\rm pole}$ and $\Gamma_{\rm pole}$  to characterize the mass and width of the $\zc(3885)$ peak.
Table~\ref{tbl:fits} lists the pole masses and widths for the $\pip D^0$ and $\pim D^+$ tagged samples.

\begin{table}[htb]
\begin{center}
\caption{\label{tbl:fits}
The pole mass $M_{\rm pole}$ and width $\Gamma_{\rm pole}$, signal yields and fit quality ($\chi^2 $/ndf)
for the two tag samples.}
\begin{tabular}{l c c c c}
\hline  
\hline  
 Tag         & {\footnotesize $M_{\rm pole}$(MeV/$c^2$)}~&~{\footnotesize $\Gamma_{\rm pole}$(MeV)}~&~{\footnotesize $\zc$ signal~(evts)}~&~$\chi^2$/ndf \\
\hline  
$\pip D^0$       &  $3882.3 \pm 1.5$  & $ 24.6 \pm 3.3 $     & $ 502\pm 41$         &  54/54      \\
$\pim D^+$       &  $3885.5 \pm 1.5$  & $ 24.9 \pm 3.2 $     & $ 710\pm 54$         &  60/54      \\
\hline  
\hline  
\end{tabular}
\end{center}
\end{table}

Monte Carlo studies of possible sources of peaking backgrounds
in the $D\dstrbar$ mass distribution show that
processes of the type $\ee\rt D \bar{D}_{X}$, $\bar{D}_X\rt \dstrbar\pi$,
would produce a near-threshold reflection peak in the
$D \dstrbar$ mass distribution, where $D_X$ denotes a
$D^*\pi$ resonance with mass near the upper kinematic boundary.
This boundary, $\sqrt{s}-m_{D}$, is $30$~MeV/$c^2$ below the mass
of the lightest established $D^*\pi$ resonance,
the $\done(2420)$, with $M_{\done}=2421.3\pm0.6$~MeV/$c^2$ and
$\Gamma_{\done} =27.1\pm 2.7$~MeV~\cite{pdg}, which suggests that
contributions from $D \donebar(2420)$ final states, either from
$Y(4260)\rt D \donebar$ decays or non-resonant
$\ee\rt D \donebar$ production, are beyond the kinematic
reach at $\sqrt{s} = 4260$~MeV and, therefore, are small.
However, some models for the $Y(4260)$ attribute it to
a bound $D \donebar$ molecular state~\cite{zhao}, where
sub-threshold $\donebar\rt \dstrbar\pi$ decays
might be important and, possibly, produce a reflection peak
in the $D\dstrbar$ mass distribution that mimics a $\zc(3885)$ signal.

To study this possibility, we separated the events into two samples
according to $|\cos\theta_{\pi D}|>0.5$ and $|\cos\theta_{\pi D}|<0.5$,
where $\theta_{\pi D}$ is the angle between the bachelor pion and the
$D$ meson directions in the $\zc(3885)$ rest frame. The $D\donebar$ MC events
predominantly have $|\cos\theta_{\pi D}|>0.5$ while, in contrast,
$\ee\rt\pi\zc$ signal-MC sample has similar numbers of events
with $|\cos\theta_{\pi D}|>0.5$ and $|\cos\theta_{\pi D}|<0.5$.
We define an asymmetry parameter ${\mathcal A}=(n_{>0.5}-n_{<0.5})/(n_{>0.5}+n_{<0.5})$,
where $n_{>0.5}$ ($n_{<0.5}$) is the fitted number of $\zc(3885)$ signal events for
$|\cos\theta_{\pi D}|>0.5$ ($<0.5$).  For the data, ${\mathcal A}_{\rm data}=0.12\pm 0.06$,
close to the MC value for $e^+e^-\rt\pi\zc(3885)$: ${\mathcal A}^{\pi\zc}_{\rm MC}=0.02\pm 0.02$,
and far from the MC result for the $e^+e^-\rt D\donebar$ hypothesis:
${\mathcal A}^{D\donebar}_{\rm MC}=0.43\pm 0.04$. We conclude that the
$D\donebar$ contribution to our observed $\zc(3885)\rt D\dstrbar$ signal
is small.

\begin{figure}[htb]
\begin{minipage}[t]{75mm}
  \includegraphics[height=0.6\textwidth,width=0.8\textwidth]{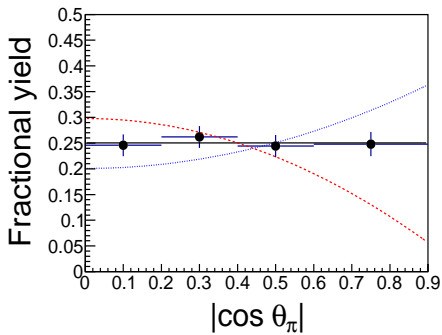}
\end{minipage}
\caption{\footnotesize $(1/N_{tot})dN/d|\cos\theta_{\pi}|~versus~|\cos\theta_{\pi}|$ for $\zc(3885)$
events in data.  The solid, dashed and dotted curves show expectations for $J^P = 1^+$, $0^-$
and $1^-$, respectively.
}
\label{fig:dndcos-pi}
\end{figure}

If the $J^P$ quantum numbers of the $\zc(3885)$ are $1^+$, the relative $\pi$-$\zc$ orbital
in the decay $Y(4260)\rt\pi\zc$ can be $S$- and/or $D$-waves. Since the
decay is near threshold, the $D$-wave contribution should be small,
in which case the $dN/d|\cos\theta_{\pi}|$ distribution would
be flat, where $\theta_{\pi}$ is the bachelor pion's polar angle relative to the beam
direction in the CM. If $J^P = 0^-$, the decay can only proceed via a
$P$-wave and is polarized with $J_z=\pm 1$; in this case
$dN/d\cos\theta_{\pi} \propto \sin^2\theta_{\pi}$.  Similarly,  $J^P=1^-$ also implies a
$P$-wave with an expected distribution that goes as $1 + \cos^2\theta$.
Parity conservation excludes $J^P=0^+$.

We sliced the data into four $|\cos\theta_{\pi}|$ bins and repeated the fits described
above for each bin. The $|\cos\theta_{\pi}|$-dependence of the efficiency is determined
from signal MC event samples.
Figure~\ref{fig:dndcos-pi} shows the efficiency-corrected fractional signal yield
{\it vs.} $|\cos\theta_{\pi}|$. The solid (dashed) curve shows the result of a fit to
a flat ($\sin^2\theta_{\pi}$) distribution. The data agree well with the flat
expectation for $J^P=1^+$, with $\chi^2 /{\rm ndf} = 0.44/3$ and disagree with those
for $J^P=0^-$, for which $\chi^2/{\rm ndf}=32/3$, and $1^-$, where $\chi^2/{\rm ndf}=16/3$.

We use the fitted numbers of signal events for the $\pip\dz$-tagged sample,
$N_{\pip}(\zc^{-}\rt (D\dstrbar)^{-})$, and for the $\pim\dplus$-tagged sample,
$N_{\pim}(\zc^{+}\rt (D\dstrbar)^{+})$ to make two independent measurements of
the product of the cross section and branching fraction
$\sigma(\ee\rt\pi \zc)\times {\mathcal B}(\zc \rt D\dstrbar)$.
We assume isospin symmetry and, for
the $\pip\dz$-tagged channel, use the relation
\begin{eqnarray}
&& \sigma(\ee\rt\pip \zc(3885)^{-})\times {\mathcal B}(\zc^{-}\rt (D\dstrbar)^-)\\
\nonumber
&=& \frac {N_{\pip}^{-}(\zc^-\rt (D\dstrbar)^{-})}
{{\mathcal L}(1+\delta){\mathcal B}_{\dz\rt K^-\pip}(\epsilon^{0}_1+\epsilon^{0}_2)/2},
\end{eqnarray}
where ${\mathcal L}=525\pm 5$~pb$^{-1}$ is the integrated luminosity, $(1+\delta)=0.87\pm 0.04$
is the radiative correction factor~\cite{factor}, $\epsilon^{0}_1=0.46$ is the efficiency for
$\pip\zc^-$, $\zc^-\rt\dz\dstrm$ MC events
and $\epsilon^{0}_2=0.21$ is the efficiency for
$\pip\zc^-$, $\zc^-\rt \dminus\dstrz$, $\dstrz\rt \gamma/\piz\dz$ MC events. The resulting value is
$\sigma(\ee\rt\pip \zc^{-})\times {\mathcal B}(\zc(3885)^{-}\rt (D\dstrbar)^-)=84.6\pm 6.9$~pb,
where the error is statistical only.

For the $\pim\dplus$-tagged channel, we use
\begin{eqnarray}
&&\sigma(\ee\rt\pim \zc(3885)^{+})\times {\mathcal B}(\zc^{+}\rt (D\dstrbar)^+) \\
\nonumber
&=& \frac {N_{\pim}(\zc^{+}\rt (D\dstrbar)^{+})}
{{\mathcal L}(1+\delta){\mathcal B}_{\dplus\rt K^-\pip\pip}(\epsilon^{+}_1+
\epsilon^{+}_2{\mathcal B}_{\dstrp\rt\piz\dplus})/2},
\end{eqnarray}
where $\epsilon^{+}_1=0.34$ is the efficiency for $\pim\zc^+, \zc^+\rt \dplus\dstrzbar$ MC events
and $\epsilon^{+}_2=0.24$ is the efficiency for
$\pim\zc^+$, $\zc^+\rt \bar{D}^0 \dstrp$, $\dstrp\rt \pi^0\dplus$ MC events.
The result is
$\sigma(\ee\rt\pim \zc(3885)^{+})\times {\mathcal B}(\zc^{+}\rt (D\dstrbar)^+)=82.3\pm 6.3$~pb
(statistical error only) and in good agreement with that for the $\pip\dz$-tag sample, which
justifies our assumption of isospin invariance.

\begin{table}[htb]
\begin{center}
\caption{\label{tbl:syst}
Contributions to systematic errors on the pole mass, pole width and signal yield.
When two values are listed, the first is for $\pip\dz$ tags and the second for $\pim\dplus$ tags.}
\begin{tabular}{l c c c}
\hline 
\hline 
Source                        &~{\footnotesize $M_{\rm pole}$(MeV/$c^2$)}~&~{\footnotesize $\Gamma_{\rm pole}$(MeV)}~&~$\sigma\times {\mathcal B}$~(\%) \\
\hline 
Tracking \& PID             &              &                &      $\pm 4/6$        \\
$D$ mass req.                &              &                &      $\pm 1$          \\
$\dz/\dplus$ Bfs.            &              &                &      $\pm 1$         \\
Kinematic fit                 &              &                &      $\pm 4$          \\
Signal BW shape               & $\pm 1/2$    &   $\pm  3$     &      $\pm 5$      \\
Bkgd shape                    & $\pm 4.0/3.8$& $\pm 10.4/10.7$&      $\pm 24$        \\
MC efficiency                 &              &                &      $\pm 6/3$        \\
$Y(4260)$ lineshape           &              &                &      $\pm 0.6$        \\
Luminosity                    &              &                &      $\pm 1$         \\
Rad. corr.                    &              &                &      $\pm 5$         \\
\hline 
Sum in quadrature             &$\pm 4.1/4.3$ &$\pm 10.8/11.1$ &      $\pm 26.4/26.3$   \\
\hline 
\hline 
\end{tabular}
\end{center}
\end{table}
Systematic errors include uncertainties from tracking, particle ID, $D$ mass and decay branching fraction,
kinematic fit, signal and background shapes, MC efficiency, $Y(4260)$ lineshape, the
radiative correction factor and the luminosity.  The uncertainties from tracking and particle
ID are both 1\% per track. The uncertainties from $D$ selection and the kinematic fit are determined from
a $e^+e^- \rt D^{*+}D^{*-}$ control sample that has the same final state as the $\pi^+D^0D^{*-}$ signal events.
The variation of the efficiency over the $\zc(3885)$ mass uncertainty range is included as a
systematic error.  The systematic errors for the luminosity and $Y(4260)$ resonance parameters
are taken from Ref.~\cite{bes3_z3900}.
For the signal shape error
we use the difference between the the pole mass \& width and signal yield
from the fits that use a mass-dependent width (default) and
the mass, width and yield from fits with mass-independent-width BW lineshapes.
The most significant contributions to the systematic errors are
related to the choice of background shape.  For this, we compare results from
the default fit with those that use a symmetric exponential threshold function
and the distribution of wrong-sign $\pi D$ events extracted from the data.

In all the fits used in this analysis, it is assumed that the $\pi \zc(3885)$
system is produced in an $S$-wave and the $D\dstrbar$ system produced in
the decay of the $\zc(3885)$ is also in an $S$-wave. Attempts to fit the peak
using $P$-wave line shapes all failed to converge. This compatibility with $S$-wave
is consistent with the observed $\cos\theta_{\pi}$ distribution.

The contributions from each source are summarized in Table~\ref{tbl:syst}.
We assume that the errors from the different sources are uncorrelated and
use the sums in quadrature as the total systematic errors.

For the final mass, width and cross section values, we use weighted averages of the results
from the two tag modes, with the near-complete correlations between the systematic errors taken
into account.  The results are listed in Table~\ref{tbl:compare}, where
we also include results for the $Z_c(3900)\rt \pi\jp$
taken from Ref.~\cite{bes3_z3900} for comparison.
When statistical and systematic errors are added in quadrature, the $\zc(3885)$ mass is
about $2\sigma$ lower than that for the $Z_c(3900)$ and the width is $1\sigma$ lower.
\begin{table}[htb]
\begin{center}
\caption{\label{tbl:compare}
Parameters for the $\zc (3885)\rt D\dstrbar$ reported here and those for the $Z_c (3900)\rt\pi\jp$
taken from Ref.~\cite{bes3_z3900}.}
\begin{tabular}{l c c}
\hline 
\hline 
                                    &~~ $\zc (3885)\rt D\dstrbar$~~&~~$Z_c(3900)\rt \pi\jp$  \\
\hline 
Mass~(MeV/$c^2$)                          &~~ $3883.9 \pm 1.5 \pm 4.2$~~&  $3899\pm 3.6 \pm 4.9$    \\
$\Gamma$~(MeV)                      & $24.8 \pm 3.3 \pm 11.0$       &  $46 \pm 10 \pm 20$        \\
$\sigma\times {\mathcal B}$~(pb) ~~& $83.5 \pm 6.6 \pm 22.0 $      &  $13.5 \pm 2.1 \pm 4.8$    \\
\hline 
\hline 
\end{tabular}
\end{center}
\end{table}

In summary, we report observation of a strong, near-threshold enhancement, $\zc (3885)$, in the
$D\dstrbar$ invariant mass distribution in the process $\ee\rt \pi^{\pm} (D\dstrbar)^{\mp}$
at $\sqrt{s}=4.26$~GeV. Attempts to fit the $\zc (3885)$ peak with a $P$-wave BW lineshape failed
to converge, and the $|\cos\theta_{\pi}|$ distribution agrees well with $S$-wave expectations; both
results favor a $J^P=1^{+}$ quantum number assignment.  Other $J\leqslant 1$ assigments are eliminated.

An important question is whether or not the source of the $\zc(3885)\rt D\bar{D}^*$
structure is the same as that for the $Z_c(3900)\rt \pi \jp$.
The fitted $\zc(3885)$ mass is about $2\sigma$ below that of the $Z_c(3900)$~\cite{bes3_z3900,belle_z3900}.
However neither fit considers the possibility of interference with a coherent non-resonant background
that could shift the results.  A $J^P$ quantum number determination of the
$Z_c(3900)^{\pm}$ would provide an additional test of this possibility.

Assuming the $\zc (3885)$ structure reported here is due
to the $Z_c(3900)$, the ratio of partial decay widths is determined to be
$\frac{\Gamma(\zc(3885)\rt D\bar{D}^*)}{\Gamma(Z_c(3900)\rt\pi\jp)}=6.2 \pm 1.1 \pm2.7$
(here the main systematic errors are almost entirely uncorrelated).
This ratio is much smaller than typical values for decays of conventional charmonium states
above the open charm threshold. For example:
$\Gamma (\psi(3770)\rt D\bar{D})/\Gamma(\psi(3770)\rt \pipi\jp)=482\pm 84$~\cite{pdg}
and $\Gamma (\psi(4040)\rt D^{(*)}\bar{D}^{(*)})/\Gamma(\psi(4040)\rt \eta\jp)=192\pm 27$~\cite{psi4040_2_etajpsi}.
This suggests the influence of very different dynamics in the $Y(4260)$-$Z_c(3900)$ system.

The BESIII collaboration thanks the staff of BEPCII and the computing center for their strong support.
 This work is supported in part by the Ministry of Science and Technology of China Contract No. 2009CB825200;
National Natural Science Foundation of China (NSFC) Contract Nos. 10821063, 10825524, 10835001, 10935007, 11125525, 11235011; Joint Funds of the National Natural Science Foundation of China under Contract Nos. 11079008, 11179007, 11079027; Chinese Academy of Sciences (CAS) Large-Scale
Scientific Facility Program; CAS Contract Nos. KJCX2-YW-N29, KJCX2-YW-N45; 100 Talents Program of CAS;
German Research Foundation (DFG) Contract No. Collaborative Research Center CRC-1044;
Istituto Nazionale di Fisica Nucleare, Italy; Ministry of Development of Turkey Contract No. DPT2006K-120470;
U. S. Department of Energy Contract Nos. DE-FG02-04ER41291, DE-FG02-05ER41374, DE-FG02-94ER40823;
U.S. National Science Foundation; University of Groningen (RuG); Helmholtzzentrum f\"{u}r
Schwerionenforschung GmbH (GSI) Darmstadt; Korean National Research Foundation (NRF) Grant No. 20110029457.

%


\begin{thebibliography}{99}

\bibitem{babar_y4260} B.~Aubert {\it et al.} (BaBar Collaboration),
Phys. Rev. Lett. {\bf 95}, 142001 (2005).

\bibitem{cleo_y4260} Q.~He {\it et al.} (CLEO Collaboration),
Phys. Rev. D {\bf 74}, 091104(R) (2006).

\bibitem{belle_y4260} C.Z.~Yuan {\it et al.} (Belle Collaboration),
Phys. Rev. Lett. {\bf 99}, 182004 (2005).

\bibitem{galina}
G.~Pakhlova {\it et al.} (Belle Collaboration),
Phys. Rev. Lett. {\bf 98}, 092001 (2007);
Phys. Rev. Lett. {\bf 100}, 062001 (2008);
Phys. Rev. D {\bf 77}, 011103 (2008);
Phys. Rev. Lett. {\bf 101}, 172001 (2008); and
Phys. Rev. D {\bf 80}, 091101 (2009).


\bibitem{mo} X.H.~Mo {\it et al.},
Phys. Lett. {\bf B640}, 182 (2006).

\bibitem{pdg} J.~Beringer {\it et al.} (Particle Data Group),
 Phys. Rev. D {\bf 86}, 010001 (2012).

\bibitem{belle_pipiyns} K.-F.~Chen {\it et al.} (Belle Collaboration),
Phys. Rev. Lett. {\bf 100}, 112001 (2008).

\bibitem{belle_z_b} A.~Bondar {\it et al.} (Belle Collaboration),
Phys. Rev. Lett. {\bf 108}, 122001 (2012).


\bibitem{belle_zb_bbstr} I.~Adachi {\it et al.} (Belle Collaboration),
arXiv:1209.6450v2 [hep-exp].

\bibitem{voloshin}
A.E.~Bondar {\it et al.}, Phys. Rev. D {\bf 84}, 054010 (2011),
D.V.~Bugg, Europhys. Lett. {\bf 96}, 11002 (2011), I.V.~Danilkin,
V.D.~Orlovsky and Yu.A.~Simonov, Phys. Rev. D {\bf 85}, 034012 (2012),
C.-Y.~Cui, Y.-L.~Liu and M.-Q.~Huang, Phys. Rev. D {\bf 85}, 054014 (2012),
T.~Guo, L.~Cao, M.-Z.~Zhou and H.~Chen, arXiv:1106.2284 [hep-ph], and
J.-R.~Zhang, M.~Zhong and M.-Q.~Huang Phys. Lett. {\bf B704}, 312 (2011).

\bibitem{molecule}See, for example,
M.B.~Voloshin and L.B.~Okun,
JETP Lett. {\bf 23}, 333 (1976);
M. Bander, G.L.~Shaw and P.~Thomas,
Phys. Rev. Lett. {\bf 36}, 695 (1977);
A.~De~Rujula, H.~Georgi and S.L.~Glashow,
Phys. Rev. Lett. {\bf 38}, 317 (1977);
A.V. Manohar and M.B. Wise, Nucl. Phys. B {\bf 339}, 17 (1993);
N.A.~T\"{o}rnqvist, hep-ph/0308277 (2003);
F.E.~Close and P.R.~Page, Phys. Lett. B {\bf 578}, 119 (2003);
C.-Y.~Wong, Phys. Rev. C {\bf 69}, 055202 (2004);
S.~Pakvasa and M.~Suzuki, Phys. Lett. B {\bf 579}, 67 (2004);
E.~Braaten and M.~Kusunoki, Phys. Rev. D {\bf 69}, 114012 (2004);
E.S.~Swanson, Phys. Lett. B {\bf 588}, 189 (2004);
D.~Gamermann and E.~Oset, Phys. Rev. D {\bf 80}, 014003 (2009)
\& Phys. Rev. D {\bf 81}, 014029 (2010).

\bibitem{bes3_z3900} M.~Ablikim {\it et al.} (BESIII Collaboration),
Phys. Rev. Lett. {\bf 110}, 252001 (2013).

\bibitem{belle_z3900}
Z.Q.~Liu {\it et al.} (Belle Collaboration), Phys. Rev. Lett. {\bf 110}, 252002 (2013).

\bibitem{zhao} Q.~Wang, C.~Hanhart and Q.~Zhao, Phys. Rev. Lett. {\bf 111}, 132003 (2013).

\bibitem{mahajan}
N.~Mahajan, arXiv:1304.1301 [hep-ph],
M.B.~Voloshin, Phys.Rev. D {\bf 87}, 091501 (2013),
J.-R.~Zhang, Phys. Rev. D {\bf 87}, 116004 (2013),
F.-K. Guo, {\it et al.}, Phys. Rev. D {\bf 88}, 054007 (2013) and
C.-Y.~Cui {\it et al.}, arXiv:1304.1850.

\bibitem{bes3_z4025} M.~Ablikim {\it et al.} (BESIII Collaboration), arXiv:1308.2760 [hep-ex]
and M.~Ablikim {\it et al.} (BESIII Collaboration), arXiv:1309.1896 [hep-ex].

\bibitem{faccini}
R.~Faccini {\it et al.}, Phys. Rev. D {\bf 87}, 111102R (2013);
and M.~Karliner and S.~Nussinov, JHEP {\bf 1307}, 153 (2013).
See, also, A.~Ali, C.~Hambrock and W.~Wang, Phys. Rev. D {\bf 85},
054011 (2012).



\bibitem{BESIII} M.~Ablikim {\it et al.} (BESIII Collaboration),
Nucl. Istrum. and Methods Phys. Res., Sect. A {\bf 614}, 345 (2010).

\bibitem{evtgen} D.J.~Lange,
Nucl. Istrum. and Methods Phys. Res., Sect. A {\bf 462}, 152 (2001).

\bibitem{kkmc} S.~Jadach, B.F.L.~Ward, and Z.~Was,
Comput. Phys.  Commun. {\bf 130}, 260 (2000);
Phys. Rev. D {\bf 63}, 113009 (2001).

\bibitem{geant4}
S.~Agostinelli~{\etal} (Geant4 Collaboration),
Nucl. Istrum. and Methods Phys. Res., Sect. A {\bf 506}, 250 (2003).

\bibitem{boost} Z.Y.~Deng, {\it et al.},
High Energy Phys. Nucl. Phys. {\bf 30} 371, (2006).

\bibitem{lund} R.G.~Ping {\it et al.}, Chinese Phys. C {\bf 32}, 599 (2008).

\bibitem{pythia} T.~Sj\"{o}strand, S.~Mrenna and P.~Skands,
JHEP {\bf 026}, 0605 (2006).

\bibitem{recoil} We minimize the effect of the $D$ mass resolution by
plotting $M^{\rm recoil}=RM(\pi D)+M(D)-m_D$, where $RM(\pi D)$
is the recoil mass inferred from four-momentum conservation and $M(D)$ is the
measured $D$ mass.


\bibitem{psi4040_2_etajpsi} M.~Ablikim {\it et al.} (BESIII Collaboration),
Phys. Rev. D {\bf 86}, 071101 (2012).


\bibitem{pole} See, for example, A.R.~Bohm and N.L.~Harshman, arXiv:hep-ph/0001206 and
references cited therein.

\bibitem{factor} We use a second-order QED calculation and unpublished
BESIII energy-dependent measurements of $\sigma(\ee\rt\pi D\bar{D^*})$
to compute the radiative correction; see E. A. Kuraev
and V. S. Fadin, Yad. Fiz. {\bf 41}, 733 (1985)
[Sov. J. Nucl. Phys. {\bf 41}, 466 (1985)].

\end{thebibliography}
\end{document}